\documentclass[sigconf]{acmart}

\usepackage{booktabs} %
\usepackage{multirow}
\usepackage{latexsym}
\usepackage{graphicx} 
\usepackage{algorithm}
\usepackage{algorithmic}
\usepackage{epsfig}
\usepackage{color}
\usepackage{amsmath}
\usepackage{graphicx}
\usepackage{makecell}
\usepackage{textcomp}
\usepackage{url}
\usepackage{flushend}
\usepackage{bbm}
\usepackage{stfloats}

\usepackage{booktabs}
\usepackage{multirow}
\usepackage{graphicx}
\usepackage{float} 
\usepackage{subfigure}
\usepackage{amsmath,amsfonts,amsthm,bm}
\usepackage{algorithmic}
\usepackage{algorithm}  
\usepackage{algorithmic,eqparbox,array}
\setcopyright{rightsretained}
\newtheorem{theorem}{Theorem}
\renewcommand{\algorithmicrequire}{\textbf{Input:}}
\renewcommand{\algorithmicensure}{\textbf{Output:}}

\begin{document}

\title{PrivateRec: Differentially Private Model Training and Online Serving for Federated News Recommendation}

\author{Ruixuan Liu$^1$, Yanlin Wang$2$, Yang Cao$^3$, Lingjuan Lyu$^4$, Weike Pan$^5$, Yun Chen$^6$, Hong Chen$^1$}
\affiliation{%
 \institution{$^1$Renmin University of China, $^2$Microsoft Research Asia, $^3$Kyoto University, $^4$Sony, $^5$Shenzhen University, $^6$Shanghai University of Finance and Economics}
 \country{}
 }
\email{
{ruixuan.liu, chong}@ruc.edu.cn,
yanlwang@microsoft.com
}
\email{
yang@i.kyoto-u.ac.jp,
lingjuan.lv@sony.com,
panweike@szu.edu.cn,
 yunchen@sufe.edu.cn }

\begin{abstract}
  Collecting and training over sensitive personal data raise severe privacy concerns in personalized recommendation systems, and federated learning can potentially alleviate the problem by training  models over decentralized user data.
  However, a theoretically private solution in both the training and serving stages of federated recommendation is essential but still lacking.
  Furthermore, naively applying differential privacy (DP) to the two stages in federated recommendation would fail to achieve a satisfactory trade-off between privacy and utility due to the high-dimensional characteristics of model gradients and hidden representations.
  In this work, we propose a federated news recommendation method for achieving a better utility in model training and online serving under a DP guarantee.
  We first clarify the DP definition over behavior data for each round in the life-circle of federated recommendation systems.
  Next, we propose a privacy-preserving online serving mechanism under this definition based on the idea of decomposing user embeddings with public basic vectors and perturbing the lower-dimensional combination coefficients. 
  We apply a random behavior padding mechanism to reduce the required noise intensity for better utility.
  Besides, we design a federated recommendation model training method, which can generate effective and public basic vectors for serving while providing DP for training participants.
  We avoid the dimension-dependent noise for large models via label permutation and differentially private attention modules.
  Experiments on real-world news recommendation datasets validate that our method achieves superior utility under a DP guarantee in both training and serving of federated news recommendations.

\end{abstract}


\maketitle
\section{Introduction}
Nowadays, news recommendation systems are indispensable for users to filter a large amount of news articles and relieve the information overload in the web applications.
To model personalized interests for accurate recommendations, most news recommendation methods~\cite{rendle2010factorization, covington2016deep, wu2019npa, an2019neural2, zhou2018deep, li2019multi} depend on training over user behaviors such as historical clicks.
During recommendation services, personal behavior data are also necessary for the model to rank candidate news articles that fit user interests.
However, the trust and privacy is also an essential concern on the wide web.
The behavior data contain abundant personal identifiable information and are risky to be exchanged directly.
Thus, an ever-increasing privacy concern arises in the whole society for the current recommendation system.
Many law regulations (e.g., GDPR\footnote{https://gdpr-info.eu}, CCPA\footnote{https://oag.ca.gov/privacy/ccpa}) are adopted to limit the transportation and exploitation of personal data.
As a result, collecting behavior data for training and serving in recommendation systems may be forbidden in the near future, making an effective recommendation service challenging.

Federated learning (FL)~\cite{mcmahan2017communication} is proposed as a new paradigm to train models on scattered data, which mitigates the privacy concern because personal data are only kept on users' devices.
Federated recommendation systems (FRS)~\cite{muhammad2020fedfast, shin2018privacy, qi2020privacy, truex2020ldp, chen2020robust, lin2020fedrec} enable multiple users and a server collaboratively train recommendation models by exchanging model gradients instead of personal data.
However, two critical challenges hinder the implementation of federated recommendations.

First, an effective solution with theoretical privacy protection throughout the life-circle of FRS is essential but still lacking.
Existing privacy attacks indicate the possibility of inferring private information in federated training by observing local model updates or even stealing more private data by manipulating global model parameters~\cite{zhu2019deep, nasr2019comprehensive}.
Also, private data can be leaked in the prediction stage~\cite{zhang2021privacy}.
Similarly, potential adversaries~\cite{chai2020secure, gao2020privacy} in FRS can infer users' clicking histories and track their personal intents.
However, existing FRS methods do not consider the theoretical privacy guarantee~\cite{muhammad2020fedfast, ammad2019federated, liang2021fedrec++, lin2020fedrec, qi2020privacy, qi2021uni} or only consider the privacy in training stage~\cite{shin2018privacy, truex2020ldp, li2021federatedMF}.
The connection between privacy-preserving training and serving has hitherto received scant attention.

Second, a utility hurdle arises in the attempt of providing a theoretical privacy guarantee for federated news recommendations.
As a golden standard of privacy criterion, local differential privacy (LDP) \cite{dwork2014algorithmic} can be applied in federated training by perturbing local model updates before sending to the server\cite{qi2020privacy, qi2021uni}.
Since the noise magnitude is dimension-dependent and news recommendation models are typically large deep models, the utility drop is significant.
In addition, for the FRS serving with millions of candidate news, it is impractical to conduct local serving~\cite{huang2020federated, chen2020practical} where each user is required to store all candidate items for local ranking due to enormous communication and memory costs.
Thus, we follow a more practical way of online serving~\cite{qi2021uni}, where user embeddings encoded with local historical clicks are sent to the server for recommendations.
Typically, a larger-dimensional user embedding is more potent to describe user interests, but the utility can be ruined with the dimension-dependent noise for a DP guarantee~\cite{dwork2014algorithmic}.
Therefore, it is challenging to provide a satisfactory utility under a reasonable privacy for both training and serving.

To solve above challenges, we propose a federated news recommendation framework \textit{PrivateRec} with a theoretical privacy guarantee and a decent recommendation performance.
First of all, we rethink the essential question of how differential privacy~\cite{dwork2014algorithmic} is defined in training and serving for federated news recommendation.
Then, we preserve the privacy of historical clicks by injecting DP noise when encoding user embeddings during federated training and serving.
Additionally, we permute the true labels in the training data for preserving users' true responses to candidate items, which results in the overall privacy guarantee of \textit{PrivateRec}.
To avoid the severe utility drop caused by dimension-dependent noise in training\cite{qi2020privacy, qi2021uni}, we decompose a user embedding into low-dimensional attentions before perturbation in both training and serving.
We further reduce the noise magnitude with the privacy amplification effect of the random padding in the user encoder.
Experiments validate a decent utility-privacy trade-off of our method and emphasize the necessity of considering the privacy-preserving serving utility when design federated recommendation solutions.

To summarize, the main contributions of this paper are three folds: 
(1) We propose a novel and unified framework \textit{PrivateRec} for better utility in training and serving of a theoretically private federated news recommendation system.
(2) We design privacy-preserving mechanisms for federated training and serving with a decent recommendation performance.
\textit{PrivateRec} is based on the idea of decomposing the user embedding into a lower-dimensional attention vector to avoid the utility drop by the dimension curse for both training and inference.
Then we design a random padding mechanism to further reduce injected noise and improve the performance.
(3) Extensive experiments and analysis on real-world datasets validate the significant utility improvement of training and serving in \textit{PrivateRec}. The utility lower bound under an extremely strong privacy is improved by 5-11\% on AUC.

\vspace{-10pt}
\section{Related Work}
Personalized news recommendation is important for intelligent online news service.
Many deep learning-based recommendation models have been proposed for this task~\cite{ wu2019npa, okura2017embedding, wu2019neural, wu2019neural2}.
Generally, their frameworks include three core components, i.e., news model, user model, and click prediction module.
The news model can learn news representations from news content with CNN~\cite{wu2019neural} or pre-trained language models~\cite{wu2021empowering}.
The user model aims to model user interest from their clicked news with components as GRU network~\cite{okura2017embedding}, personalized attention~\cite{wu2019npa} and multi-head self-attention~\cite{wu2019neural2}.
Last, the clicking prediction module estimates the matching score between a news embedding and a user embedding, which can be implemented with the dot product~\cite{an2019neural}, the outer product~\cite{he2018outer} and the dense network~\cite{wang2018dkn}.
However, these works rely on centralized user data for model training and serving and violates privacy~
\cite{qi2020privacy}.

Federated recommendation systems (FRS)~\cite{ammad2019federated, lin2020fedrec, huang2020federated, qi2020privacy, wu2021fedgnn}  are adapted from FedAvg~\cite{mcmahan2017communication} to avoid centralized data collection.
Since exchanging gradients still poses a threat to user privacy \cite{chai2020secure, zhu2019deep}, privacy-preserving FRS methods are necessary.
Homomorphic encryption can be used to encrypt local model update~\cite{bonawitz2017practical, chai2020secure}, but it largely increases the communication or computation cost for deep models.
Anonymity~\cite{lin2020fedrec} or perturbation~\cite{qi2020privacy} can mitigate the privacy risks, but do not provide theoretical privacy guarantee as DP.
A bandit framework~\cite{li2020federated} for FRS apply DP to output a private sum reward, but cannot be generalized to recommendation model training.
A comprehensive study~\cite{li2021federatedMF} discussed vertical, horizontal and local FRS with per-rating and per-user privacy DP, but only limits to matrix factorization model.
It should be noted that existing works for FRS are designed under different DP definitions, such as bounding the influence of a local App \cite{huang2020federated}, the meta feature for an item \cite{chen2020practical}, or the indistinguishment for any two users \cite{shin2018privacy, nguyen2016collecting}.
There also multiple DP definitions applied in general FL, such as user-level~\cite{mcmahan2017learning}, sample-level~\cite{abadi2016deep}, local-user-level~\cite{nguyen2016collecting} and parameter-level~\cite{shokri2015privacy}.
As far as we know, there is no investigation for privacy definitions in deep recommendation models.

Moreover, only a few works~\cite{qi2021uni, huang2020federated, chen2020practical} discuss the model serving in federated learning.
A unified framework~\cite{qi2021uni} is proposed for training and serving with both recall and ranking models, while our work is built for ranking models.
Even the noise is injected on gradients for training and on user interests for serving, there is no theoretical privacy definition.
Different from ours, they perturb the gradient vectors in training which result in a model-size-dependent noise and cannot reach a descent utility-privacy trade-off if we formulate their noise as to a DP level.
In addition, each user in Uni-FedRec~\cite{qi2021uni} sends multiple interest vectors, which enlarges the privacy risks by multiple times given a same perturbation level for each vector.
In other works \cite{huang2020federated, chen2020practical}, the server sends all items to a user to perform local serving with the local data on multiple platforms, which does not violate privacy but is impractical for the high communication and memory cost.
In this work, we fill the gap by revisiting various DP definitions and formulating a privacy notion for every communication throughout the life-circle of federated recommendation.

\section{Methodology}
\subsection{Threat Model and Privacy Definition}
There are two parties in \textit{PrivateRec}:
(1) $\mathcal{S}$, the server who organizes the model training and returns the recommended results in serving.
Typically, $\mathcal{S}$ is an honest-but-curious server that follows the standard workflow but is curious about user's behavior data for more commercial interests.
(2) $\mathcal{U}= \mathcal{U}_{t} \cup \mathcal{U}_{s}$, the set of users $\mathcal{U}_{t}$ who participant in federated training and the set of users $\mathcal{U}_{s}$ who query recommendation services.
Specifically, federated news recommendation requires local historical clicking log $N_h=\{n_1, \cdots, n_H\}$ and clicked candidate clicking log $N_c=\{n_1, \cdots, n_C\}$ for training, and need $N_h$ to encode user interests for serving.
The information that any potential third-party adversary  can access is no more than $\mathcal{S}$, so \textit{PrivateRec} aims to protect local $N_h$ and $N_c$ for each user against an untrusted server $\mathcal{S}$.

We revisit standard DP definitions in FL as follows:
(1) User-level~\cite{mcmahan2017learning}: the adversary cannot identify the existence of a user by observing the distributed model parameters, which requires the server to be trusted.
Complementing it with SecureAgg~\cite{bonawitz2017practical} can defend untrusted servers but requires a non-negligible cost for training and cannot be applied in the serving stage. 
(2) Sample-level~\cite{abadi2016deep}: when DPSGD~\cite{abadi2016deep} is conducted in local optimization, the adversary cannot identify the existence of a training sample in the local dataset by observing the uploaded updates.
This DP definition requires the assumption that each sample is independently sampled from a distribution.
However, the training data in news recommendation models typically include samples with partially overlapping historical clicks and do not satisfy this assumption.
Also, DPSGD cannot be applied in model serving.
(3) Local-user-level~\cite{nguyen2016collecting}: any party cannot distinguish the local updates from any two different users.
Since effective news recommendation models in the industry are typically deep models~\cite{wu2021empowering}, applying local DP on local updates suffers from the notorious utility drop issue.
(4) Parameter-level~\cite{shokri2015privacy}: any party that observes a value of local update cannot identify the existence of a local training sample, which is a relaxed definition of sample-level DP  and faces the same problems for training and serving.

In \textit{PrivateRec}, we define the differential privacy as the plausibility of whether an item is clicked by a user against $\mathcal{S}$, which is derived from the central model for DP in a local view as Definition \ref{def_dp}.
This fine-grained DP definition would bring advantages of:
(1) unified privacy standard across training and serving, which enable users to quantify the overall privacy cost in a federated recommendation service with a clear notion.
(2) support for personalized privacy settings to define which news categories are non-sensitive (e.g., entertainment) and spend privacy budget only for sensitive ones (e.g., politics, health).

\begin{figure*}[!t]
    \centering
    \includegraphics[trim=80 200 80 180, clip, width=\textwidth]{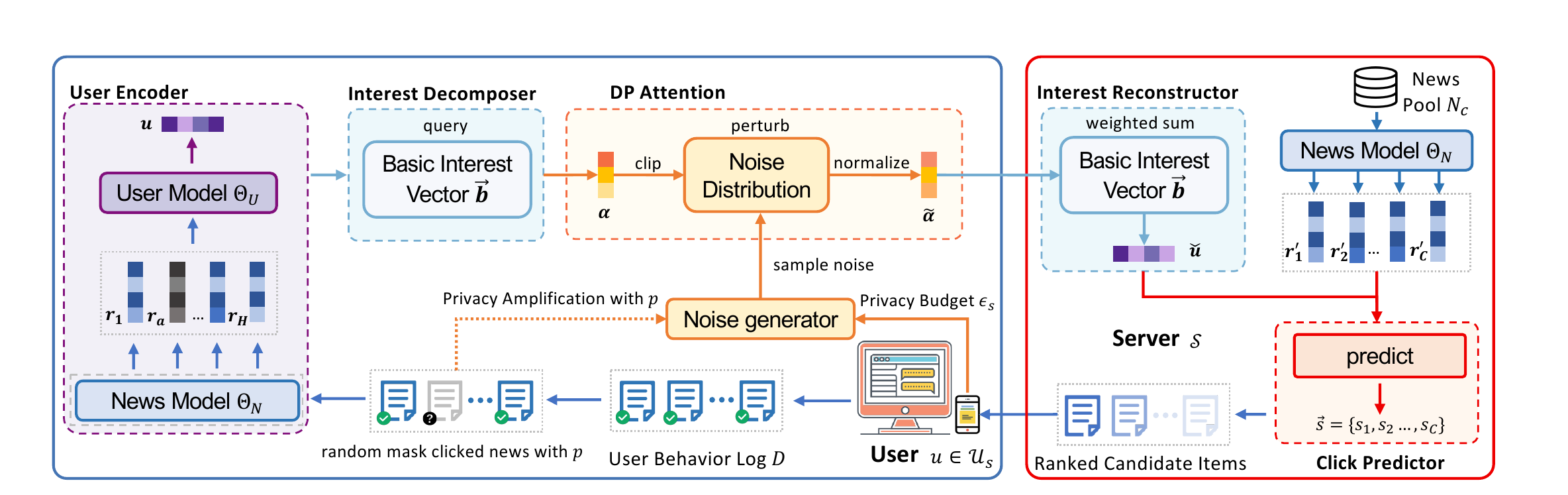}
    \caption{Privacy-preserving online serving in \textit{PrivateRec}.}
    \label{fig-overview-serve}
\end{figure*}

\begin{figure*}[t]
    \centering
    \includegraphics[trim=30 120 30 165,clip,width=0.99\textwidth]{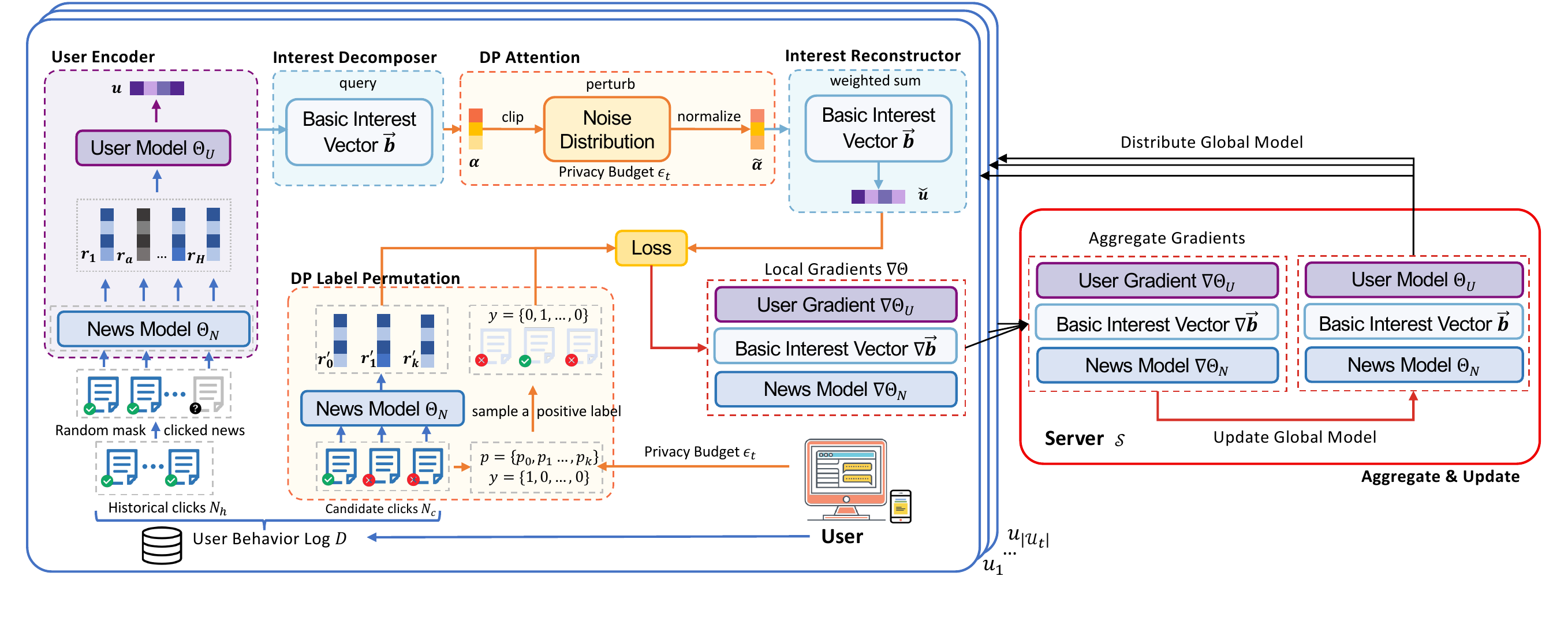}
    \caption{Privacy-preserving model training in \textit{PrivateRec}.}
    \label{fig-overview-train}
\end{figure*}

\vspace{-10pt}
\subsection{Privacy-Preserving Online Serving}
In this section, we introduce detailed modules in \textit{PrivateRec} serving as shown in Fig.~\ref{fig-overview-serve}, which lay the foundation for designing \textit{PrivateRec} training.
When the training finishes, any user with a deployed model can request for a recommendation service via online serving by sending the user embedding $\mathbf{u}$ to $\mathcal{S}$.
With mechanism $\mathcal{M}$ on the user-end and a news pool $N_c$ on the server-end, we are supposed to ensure $\mathcal{M}$ satisfies $(\epsilon_s, \delta)$-DP in Definition \ref{def_dp} when a user sends $\mathcal{M}(D)$ to the server for recommendation results over $N_c$.
\begin{definition}\label{def_dp}
	For two adjacent user behavior logs $D \in\mathbb{D}$ and $D^\prime \in\mathbb{D}$ with only one clicked behavior different and any output $z \in \text{Range}(\mathcal{M})$, a mechanism $\mathcal{M}: D\rightarrow \text{Range}(\mathcal{M})$ is $(\epsilon, \delta)$-differentially private if and only if:
	$$
	\Pr[\mathcal{M}(D) = z] \le e^\epsilon \cdot  \Pr[\mathcal{M}(D^\prime) = z] + \delta.
	$$
\end{definition}

\begin{algorithm}[h]
	\small
	\caption{Online Serving in \textit{PrivateRec} }
	\renewcommand{\algorithmicrequire}{\textbf{User: $u \in \mathcal{U}_{s}$: }}
    \renewcommand{\algorithmicensure}{\textbf{Server $\mathcal{S}$: }}
	\begin{algorithmic}[1]\label{alg-serving}
		\REQUIRE 
		$N_h,
		\theta,
		\epsilon_s,
		\delta_s,
		p,
		\Theta = 
		\overrightarrow{\mathbf{b}} \circ
		\Theta_U \circ
		\Theta_N$
		\STATE $\tilde{\bm{\alpha}} = \text{GetPrivAttn}(
		N_h, 
		\theta, 
		\epsilon_s,
		\delta_s,
		p, 
		\overrightarrow{\mathbf{b}},
        \Theta_U, 
        \Theta_N
		)$
        \STATE send $\tilde{\bm{\alpha}}$ to $\mathcal{S}$
		\ENSURE $\tilde{\bm{\alpha}}, \overrightarrow{\mathbf{b}}, N_s$
		\STATE reconstruct user embedding $\check{\mathbf{u}} = \sum_i^B \tilde{\alpha_i}\mathbf{b}_i$ \label{line-IR}
		\STATE get news embeddings for $n_i \in N_c$ with $\mathbf{r}_{i\in[C]}^\prime=\Theta_N(n_i)$
		\STATE get matching scores over $C$ news $\overrightarrow{s} = \{ s_i=\check{\mathbf{u}}^\top \cdot r_i^\prime, i \in [C]\}$
		\STATE return the ranked news list
	\end{algorithmic}
\end{algorithm}

\begin{algorithm}[t]
    \small
    \caption{GetPrivAttn($\cdot$)}
    \begin{algorithmic}[1]\label{alg-attn}
        \REQUIRE $N_h, \theta, \epsilon,
        p,
        \overrightarrow{\mathbf{b}},
        \Theta_U, \Theta_N=\Theta_e \circ \Theta_f \circ \mathbf{e}_0$
        \ENSURE $\tilde{\bm{\alpha}}$
        \STATE $\mathbf{r}_0 = \Theta_f(\overrightarrow{\mathbf{e}_0})$, sensitivity $S = \theta$ \label{line-pad-start}
        \FOR{ each clicked item $n_i$ $\in N_h$}
		    \STATE $\mathbf{r}_i =\Theta_N(n)$ \label{line-serving-flip}
		    \STATE
		    $
    		\tilde{\mathbf{r}}_i =
    		\begin{cases}
    		\mathbf{r}_i & \text{w.p. } 1-p,\\
    		\mathbf{r}_0 & \text{w.p. } p.
    		\end{cases}
    		$
		\ENDFOR \label{line-pad-end}
		\STATE user embedding $\mathbf{u}= \Theta_U(\overrightarrow{\tilde{\mathbf{r}}})$
        \STATE attention score $\bm{\alpha}= \text{Softmax}(\frac{QK^\top}{\sqrt{d}})$, where $Q=\mathbf{u}, K=\overrightarrow{\mathbf{b}}$  \label{line-dec} 
        \STATE clipping $\bar{\bm{\alpha}} = \frac{\bm{\alpha}}{\max (1, \frac{||\bm{\alpha}||_2}{\theta})}$
        \STATE sample noise vector $\tilde{n} \sim \mathcal{N}(0, \sigma^2 \textbf{I}_{B\times B}), \sigma =\frac{S}{\log \frac{e^{\epsilon} - p}{1-p}}\sqrt{2\log\frac{1.25(1-p)}{\delta}}$ \label{line-pert-start}
        \STATE return private attention vector  $\tilde{\bm{\alpha}} = \{ \tilde{\bm{\alpha}}_i = \frac{\text{SoftPlus}(\bar{\alpha}_i + \tilde{n}_i)}{\sum_j \text{SoftPlus} (\bar{\alpha}_j + \tilde{n}_j)}
        , i\in[B]
        \}$ \label{line-pert-end}
    \end{algorithmic}
\end{algorithm}

\subsubsection{Vanilla DP User Embedding (VDP)}
A conventional paradigm \cite{dwork2006calibrating} for this goal is locally perturbing the user embedding.
The user embedding is first clipped to $\theta$ with $\bar{\mathbf{u}}  = \frac{\mathbf{u}}{\max{(1, \frac{||\mathbf{u}||_2}{\theta})}}$.
Then a noise vector $\tilde{n}$ is drawn from the Gaussian distribution $\mathcal{N}(0, \sigma^2 \textbf{I}_{d\times d})$, where $\sigma=\frac{S}{\epsilon_s}\sqrt{2\log\frac{1.25(1-p)}{\delta}}$.
The sensitivity $S = \max_{D\simeq D^\prime} ||\mathcal{M}(D)-\mathcal{M}(D^\prime)||_2 = 2\theta$ bounds the maximum change that one clicked item causes to the user embedding.
Finally, the user sends $\tilde{\mathbf{u}}= \bar{\mathbf{u}} + \tilde{\bm{n}}$ to the server.
If the Laplace mechanism is applied when $\delta=0$, $\tilde{\mathbf{n}}$ is drawn from $\text{Lap}(S/\epsilon_s)$.
Usually, the user embedding with a larger dimension encodes more information. 
However, it is obvious that the intensity of noise $\mathbb{E}[||\tilde{\bm{n}}||^2]$ scales with the dimension $d$ of the user embedding.
In other words, the information in a higher-dimensional user embedding would be submerged by the noise.
Hence, VDP cannot provide a decent trade-off between utility and privacy, as we validate in Section \ref{sec-exp}.

\subsubsection{DP User Embedding with Interest Decomposition}
Thus, we are motivated to reduce the intensity of noise by decomposing $\mathbf{u}$ into a lower-dimensional vector before the perturbation.
As shown in Algorithm \ref{alg-serving}, each user in \textit{PrivateRec} sends the perturbed lower dimensional vector, which is motivated by the interest decomposition idea of Uni-FedRec\cite{qi2021uni}.
Different from Uni-FedRec\cite{qi2021uni}, we decompose a single user embedding instead of a set of interest embeddings for multiple clusters.
The server can reconstruct a user embedding to the original dimension $d$ with $\overrightarrow{\mathbf{b}}$, which is a set of basic vectors to represent $B$ abstract user interests. 
$\overrightarrow{\mathbf{b}}$ is public for $\mathcal{U}_s$ and privacy-preserving for $\mathcal{U}_t$ because training it has spent the privacy budget $(\epsilon_t, \delta_t)$ of $\mathcal{U}_t$.

The key step of the interest decomposition is Line \ref{line-dec} in Algorithm \ref{alg-attn}.
We decompose the user embedding $\mathbf{u}$ into a low-dimensional vector $\bm{\alpha}$
by querying $\mathbf{u}$ to  $\overrightarrow{\mathbf{b}}$ with a scaled dot product.
Since variant user interests can be generalized into several basic vectors \cite{wu2021two}, we have $B \ll d$.
Thus, the intensity of the DP noise only scales with $B$, which avoids the dimension curse on $\mathbf{u}$.

Then, we provide the DP guarantee by perturbing the vector $\bm{\alpha}$ with the \underline{D}ifferentially \underline{P}rivate \underline{A}ttention (DPA) module, as shown from Line \ref{line-pert-start} to Line \ref{line-pert-end} of Algorithm \ref{alg-attn}.
It should be noted that the sensitivity is $S=\theta$ because the attention scores are always positive for any two adjacent datasets.
We use SoftPlus \cite{glorot2011deep} for positive attention weights and apply normalization to keep the summation to 1.
After receiving a $B$ dimensional attention vector $\bm{\alpha}$, $\mathcal{S}$ can reconstruct the user embedding with a weighted summation over $\overrightarrow{\mathbf{b}}$ in Line \ref{line-IR} of Algorithm \ref{alg-serving}.
So the communication cost that a user spends for a recommendation query is reduced from $O(d)$ to $O(B)$.

\subsubsection{Privacy Amplification by Behavior Padding}
Based on the interest decomposition, we further reduce the noise magnitude with the privacy amplification effect from a random padding, as shown from Line \ref{line-pad-start} to Line \ref{line-pad-end} in Algorithm \ref{alg-attn}.
In other words, by randomly masking some items with public information, we can apply a smaller $\sigma$ for a given privacy.
\begin{figure}[htb]
    \centering
    \includegraphics[trim=180 250 370 210,clip,width=0.4\textwidth]{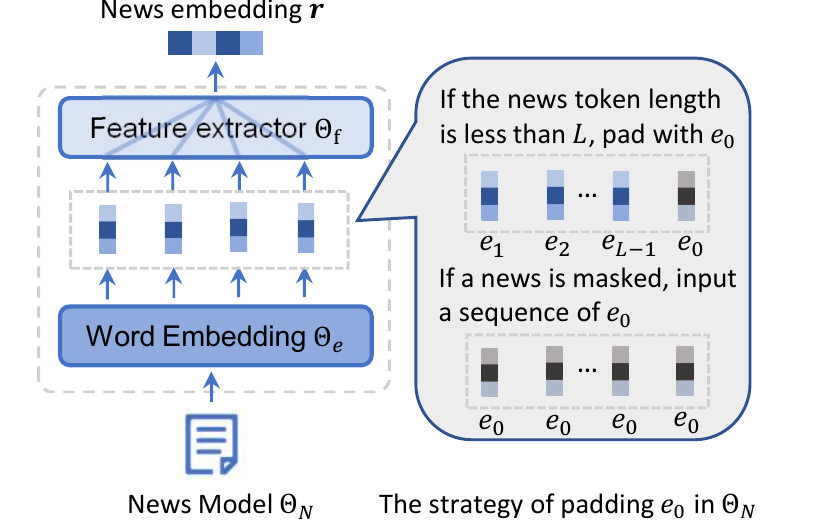}
    \caption{Details of the news model and the padding vector.}
    \label{fig-overview-pad}
    \vspace{-0.2in}
\end{figure}
Previous works \cite{wang2018not, lyu2020differentially} that utilize this effect usually mask values into null, which inevitably incur the information loss.
Instead, we pad the masked news to an anonymous news embedding $\bm{r}_0$ with the probability of $p$.
It is generated from the feature extractor  $\Theta_f$ by inputting a sequence of padding token embedding $\mathbf{e}_0$, as shown in Fig.~\ref{fig-overview-pad}.
Since the padding token embedding $\mathbf{e}_0$ is used to pad empty token in training, the generated $\bm{r}_0$ is more informative than null because it can encode some general interest information.

Same as $\overrightarrow{\mathbf{b}}$, $\mathbf{e}_0$ is trained by spending $\mathcal{U}_t$'s privacy budget, thus padding with the generated $r_0$ does not incur any extra privacy concerns than masking items to null values for $\mathcal{U}_s$.
It can improve the model serving utility with less noise under the same level of privacy while reducing the information loss caused by masking actual items to null.
We will elaborate on how to train the padding token embedding $\mathbf{e}_0$ and basic vectors $\overrightarrow{\mathbf{b}}$ in the next section.
Ultimately, we can derive the privacy guarantee as follows.
\begin{theorem}\label{theo-dp-serving}
	With GetPrivAttn($\cdot$) as the local mechanism $\mathcal{M}$, sending $\tilde{\bm{\alpha}}=\mathcal{M}(D)$ to $\mathcal{S}$ in Algorithm \ref{alg-serving} is $(\epsilon_s, \delta_s)$-DP for any $D\in \mathbb{D}$.
\end{theorem}

\vspace{-10pt}
\subsection{Privacy-Preserving Model Training}
As shown in Fig.~\ref{fig-overview-train},
the global model in \textit{PrivateRec} is $\Theta=\Theta_{N}\circ \Theta_{U} \circ \overrightarrow{\mathbf{b}}$.
Compared to the conventional federated news recommendation framework \cite{qi2020privacy}, we introduce a set of basic vectors $\overrightarrow{\mathbf{b}}$ for getting the basics to decompose the user embedding.
For each training round as shown in Algorithm \ref{alg-training}, the server $\mathcal{S}$ first samples $r$ percent users from $\mathcal{U}_{t}$ and distributes $\Theta$ to them.
Then, the sampled user $u_i$ updates the local copy of $\Theta$ with the mechanism $\mathcal{M}$ and sends $\triangledown \Theta_i$ to the server.
Finally, the server aggregates local gradients and updates $\Theta$ with FedAdam \cite{reddi2020adaptive}.
We aim to guarantee the $\mathcal{M}$ that outputs local gradients satisfies $(\epsilon_t, \delta_t)$-DP over $D=N_h\circ N_c$.

The DPA module from Line \ref{line-train-dpa} provides $(\epsilon_t, \delta_t)$-DP for $N_h$.
Recall that we require an anonymous news embedding $\mathbf{r}_0$ which can encode general context information for all items.
We illustrate two kinds of padding in Fig.~\ref{fig-overview-pad}.
For a news of which the token length is less than $L$, we pad it with $\bm{e}_0$.
For a news that is sampled to be masked, we replace all word embeddings with $\bm{e}_0$ for getting a anonymous news embedding $\bm{r}_0$
Hence, as the news model $\Theta_N$ learning to capture the context information, the padding vector $\mathbf{e}_0$ is trained to encode general information across all items, which makes it more informative than a direct nullification \cite{wang2018not}.
With the post-processing property \cite{dwork2014algorithmic}, the privacy guarantee holds for $N_h$ for the rest local processing.
For a more stable model convergence, we replace the SoftPlus function \cite{glorot2011deep} in Algorithm \ref{alg-attn} to Relu \cite{agarap2018deep}.

The \underline{D}ifferentially \underline{P}rivate \underline{L}abel Permutation (DPL) module is designed for protecting clicking behavior in $N_c$.
For each item in the candidate set $N_c$, we privately permute the label by sampling a positive one based on the exponential mechanism \cite{mcsherry2007mechanism}, as shown in Line \ref{line-exp} of Algorithm \ref{alg-training}.
With the parallel composition property of differential privacy \cite{li2016differential}, we have:
\begin{theorem}\label{theo-dp-training}
    With LocalUpdate($\cdot$) as the local mechanism $\mathcal{M}$, sending $\triangledown \Theta=\mathcal{M}(D)$ to $\mathcal{S}$ in Algorithm \ref{alg-training} is $(\epsilon_t, \delta_t)$-DP for any $D \in \mathbb{D}$.
\end{theorem}

\begin{algorithm}[t]
	\small
	\caption{ Model Training in \textit{PrivateRec}} \label{alg-training}
	\begin{algorithmic}[1]\label{alg-training}
		\REQUIRE 
		$\Theta=\Theta_N \circ \Theta_U \circ \overrightarrow{\mathbf{b}} , \epsilon_t, \delta_t, D=N_h\circ N_c$, r
		\ENSURE $\Theta^T$
		\STATE $\triangleright  \text{ Run by the server}  \mathcal{S}$
		\STATE initialize $\Theta$
		\FOR {round $t \in [T]$}
			\STATE sample $m$ users from $ \mathcal{U}_{t}$, $m= \lfloor r \cdot |\mathcal{U}_t| \rfloor $
			\FOR {each sampled user $u_i$}
			    \STATE pull $\Theta^t$ from the $\mathcal{S}$
				\STATE $\triangledown \Theta_i^t = $ LocalUpdate($\Theta^t$)
			\ENDFOR
			\STATE $\Theta^{t+1} = FedAdam(\Theta^{t}, \triangledown \Theta^t_{i\in[m]})$ 
		\ENDFOR
		\STATE deploy the final global model $\Theta^T$ to local devices
		\STATE
		\STATE $\triangleright$ LocalUpdate($\cdot$)
		\STATE $\tilde{\bm{\alpha}} = \text{GetPrivAttn}(N_h, \theta,
		\epsilon_t,
		\delta_t,
		p,
		\overrightarrow{\mathbf{b}},
		\Theta_U,
		\Theta_N
		)$ \label{line-train-dpa}
		\STATE reconstruct user embedding $\check{\mathbf{u}} = \sum_i^B \tilde{\alpha_i}\mathbf{b}_i$
		\STATE randomly sample $k$ non-clicked items from $N_c$
		\STATE sampling probability $p_{i\in[k+1]}=\frac{e^{y_i\beta}}{k+e^\beta}$ where $\beta=\max \{0, \log \frac{k}{C-1} e^{\epsilon_t}\}$ \label{line-exp}
		\STATE sample one item with the probabilities of $p_{i\in[k+1]}$ and set $y_i=1$ 
		\STATE get candidate news embeddings for $n_i \in N_c$ with $\mathbf{r}_{i\in[k+1]}^\prime=\Theta_N(n_i)$
		\STATE calculate matching score for each item  $s_{i\in[k+1]}=\mathbf{u}^\top \mathbf{r}_i^\prime$
		\STATE get the gradient $\triangledown \Theta = \frac{\partial \mathcal{L}}{\partial \Theta}$, where $\mathcal{L}=-\sum_{i=1}^{k+1} y_i \times \log \frac{e^{s_i}}{\sum_{j=1}^{k+1} e^{s_j}}$
	\end{algorithmic}
\end{algorithm}

\vspace{-10pt}
\section{Experiments}\label{sec-exp}
We conduct experiments on two real-world datasets: MIND\footnote[1]{MIND-small from \url{https://msnews.github.io/}} \cite{wu2020mind} and NewsFeeds with three news recommendation models: NRMS \cite{wu2019neural2}, NAML \cite{wu2019neural}, and PLM-NR \cite{wu2021empowering}.
Baselines in our experiments include: 
1) \textit{Centralized} recommendation, where the server trains the model over all collected personal data. 
2) \textit{DP-FedRec}, where a global recommendation model is trained over local data. Local gradient and user embedding is perturbed for training and serving in \textit{DP-RedRec} for a comparison under the same privacy level. 
Under the privacy definition \ref{def_dp}, the noise is applied to each local gradient in training with gradient clipping norm $\theta=0.005$ and to the user embedding in serving with clipping threshold $\theta=0.001$.
It should be noted that $\epsilon=\infty$ indicates the non-private \textit{FedRec} without any noise.
We show results with the off-the-shelf Laplacian mechanism\cite{dwork2014algorithmic} and $\delta=0$.
Gaussian mechanism\cite{dwork2014algorithmic, balle2018improving} can achieve a similar trend.
Details of the baselines and hyperparameters are listed in the Appendix.

\subsection{Private Training Performance}

\begin{table*}[t]
\centering
\setlength{\tabcolsep}{3.5pt}
\caption{Performance of centralized and federated training methods. ($\epsilon_t$ and $\epsilon_s$ indicate the privacy budget for training and serving under the Definition \ref{def_dp}. $\infty$ indicates no privacy protection is applied.
- indicates direct personal data collection.)}
\label{tab-train}
\begin{center}
\begin{tabular}{ccccccc|ccccc}
\toprule[1pt]
\textbf{Model}  & $\mathbf{\epsilon_t}$ & \textbf{Baseline}   & \multicolumn{4}{c}{\textbf{NewsFeeds}}                     & \multicolumn{4}{c}{\textbf{MIND}}                                                            \\ \cline{4-11}
      &                  &            & AUC                 & MRR                 & nDCG5               & nDCG10              & AUC                 & MRR                 & nDCG5               & nDCG10              \\
\hline
NAML   & -      & Centralized & 64.01$\pm$0.14 & 31.08$\pm$0.13 & 33.31$\pm$0.14 & 41.15$\pm$0.14 & 65.35$\pm$0.16 & 31.36$\pm$0.07 & 33.82$\pm$0.09 & 39.55$\pm$0.08 \\
      & $\infty$        & DP-FedRec  & 63.22$\pm$0.08          & 30.29$\pm$0.08          & 32.34$\pm$0.09          & 40.27$\pm$0.08          & 63.04$\pm$0.10          & 29.93$\pm$0.08          & 32.08$\pm$0.10          & 37.75$\pm$0.10          \\
      & 10             & DP-FedRec  & 54.84$\pm$0.55          & 25.08$\pm$0.34          & 25.67$\pm$0.43          & 33.47$\pm$0.44          & 55.31$\pm$0.31          & 24.88$\pm$0.16          & 25.91$\pm$0.16          & 31.52$\pm$0.17          \\
      & 10              & PrivateRec & 61.05$\pm$0.37 & 28.46$\pm$0.32 & 30.03$\pm$0.38 & 38.12$\pm$0.37 & 57.51$\pm$0.42 & 26.27$\pm$0.26 & 27.76$\pm$0.31 & 33.34$\pm$0.26 \\
\hline
NRMS   & -   & Centralized & 64.48$\pm$0.13 & 31.57$\pm$0.12 & 33.94$\pm$0.14 & 41.72$\pm$0.14 & 65.82$\pm$0.11 & 31.68$\pm$0.08 & 34.29$\pm$0.06 & 39.99$\pm$0.07 \\
      & $\infty$        & DP-FedRec  & 63.32$\pm$0.19          & 30.46$\pm$0.18          & 32.56$\pm$0.21          & 40.44$\pm$0.21          & 62.80$\pm$0.47          & 29.59$\pm$0.34          & 31.66$\pm$0.37          & 37.32$\pm$0.34          \\
      & 10            & DP-FedRec  & 55.48$\pm$0.74          & 25.22$\pm$0.40          & 25.91$\pm$0.54          & 33.82$\pm$0.58          & 54.59$\pm$0.54          & 24.75$\pm$0.27          & 25.64$\pm$0.32          & 31.28$\pm$0.30          \\
      & 10             & PrivateRec & 61.69$\pm$0.16 & 29.04$\pm$0.15 & 30.70$\pm$0.23 & 38.77$\pm$0.20 & 58.32$\pm$0.20 & 26.56$\pm$0.28 & 27.93$\pm$0.69 & 33.67$\pm$0.49 \\
\hline
PLM-NR & -      & Centralized & 64.69$\pm$0.21 & 31.42$\pm$0.19 & 33.71$\pm$0.24 & 41.67$\pm$0.21 & 67.08$\pm$0.15 & 32.49$\pm$0.10 & 35.21$\pm$0.14 & 40.94$\pm$0.13 \\
      & $\infty$             & DP-FedRec  & 63.48$\pm$0.35          & 30.35$\pm$0.26          & 32.46$\pm$0.33          & 40.46$\pm$0.33          & 66.05$\pm$0.25          & 31.72$\pm$0.21          & 34.28$\pm$0.30          & 40.05$\pm$0.28          \\
      & 10             & DP-FedRec  & 57.74$\pm$0.11          & 26.05$\pm$1.17          & 27.01$\pm$1.48          & 35.19$\pm$1.41          & 54.87$\pm$0.21          & 24.45$\pm$0.25          & 25.51$\pm$0.33          & 31.18$\pm$0.25          \\
      & 10            & PrivateRec & 62.52$\pm$0.17 & 29.60$\pm$0.18 & 31.50$\pm$0.21 & 39.52$\pm$0.22 & 58.75$\pm$0.14 & 26.70$\pm$0.28 & 28.22$\pm$0.32 & 33.90$\pm$0.25 \\
\bottomrule[1pt]
\end{tabular}
\end{center}
\end{table*}

First, we focus on evaluating the utility of privacy-preserving federated training by setting $\epsilon_s=\infty$ and compare \textit{PrivateRec} with \textit{DP-FedRec}.
From Table, \ref{tab-train}, 
we can see that models in \textit{Centralized} recommendation can achieve the best performance at the cost of %
user privacy.
\textit{FedRec} with $\epsilon_t=\infty$ can mitigate the direct privacy leakage without data collection and achieve the suboptimal performance, but no theoretical privacy is ensured during training.
For privacy-preserving baselines, we observe that \textit{PrivateRec} outperforms \textit{DP-FedRec} over both datasets with higher metric scores.
Also, we observe a larger utility drop from \textit{FedRec} with $\epsilon_t=\infty$ to \textit{DP-FedRec} on the PLM-NR model, because the noise amount on gradients scales with the model size.

\begin{figure}[t]
    \centering
    \includegraphics[width=0.49\textwidth]{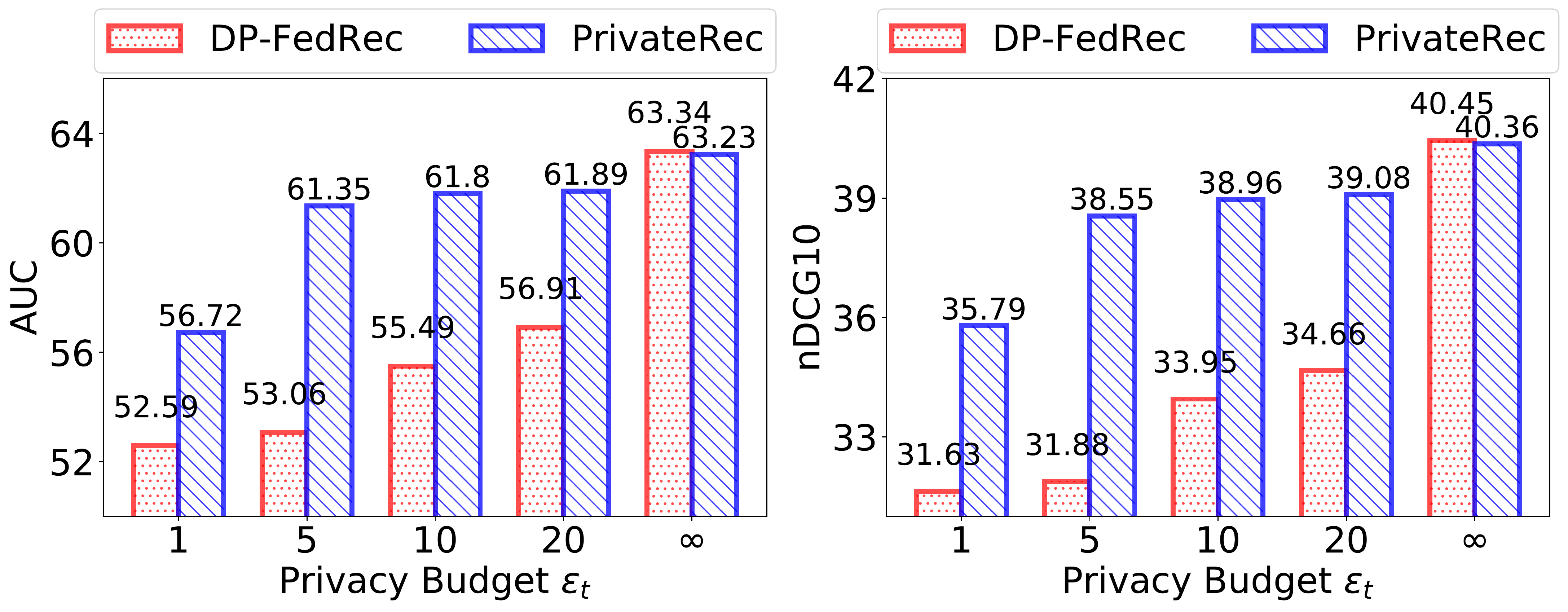}
    \caption{Performance of \textit{DP-FedRec} and \textit{PrivateRec} with different $\epsilon_t$ in federated training on NewsFeeds.}
    \label{fig-train-inf-eps}
\end{figure}

Then, we visualize the privacy-utility trade-off in federated training with various privacy budget $\epsilon_t=\{1, 5, 10, 20, \infty\}$ in Fig.~\ref{fig-train-inf-eps}.
First, the model performance of \textit{PrivateRec} is better than \textit{DP-FedRec} for all budgets $\epsilon_t$.
Additionally. the performance of \textit{PrivateRec} for a small privacy budget $\epsilon_t=5$ is still acceptable with AUC 61.35.
Second, the decomposition of user interest has a negligible effect on the federated training utility with $B=5$.
When $\epsilon_t=\infty$, the performance of \textit{PrivateRec} is comparable to the non-private baseline with $\epsilon_t=\infty$.
This observation validates our claim that thousands of user interests can be summarized into several basic interests.

\subsection{Privacy-Preserving Serving Performance} \label{sec-serve}

\begin{table*}[]
\setlength{\tabcolsep}{3.5pt}
\caption{Performance of privacy-preserving model serving.}
\label{tab-serve}
\begin{center}
\begin{tabular}{cccccccc|cccc}
\toprule[1pt]
\textbf{Model}  & $\mathbf{\epsilon_t}$ & $\mathbf{\epsilon_s}$ & \textbf{Baseline}   & \multicolumn{4}{c}{\textbf{NewsFeeds}}                     & \multicolumn{4}{c}{\textbf{MIND}}                         \\ \cline{5-12}
       &              &              &            & AUC        & MRR        & nDCG5      & nDCG10     & AUC        & MRR        & nDCG5      & nDCG10     \\
\hline
NAML   & $\infty$     & 10           & DP-FedRec  & 50.59±0.08 & 22.44±0.06 & 22.28±0.04 & 30.19±0.03 & 50.32±0.03 & 22.75±0.05 & 23.41±0.04 & 28.94±0.03 \\
       & 10           & 10           & DP-FedRec  & 50.25±0.05 & 22.21±0.06 & 21.96±0.07 & 29.92±0.03 & 50.11±0.09 & 22.69±0.07 & 23.31±0.09 & 28.83±0.08 \\
       & 10           & 10           & PrivateRec & \textbf{61.63±0.19} & \textbf{29.10±0.21} & \textbf{30.81±0.26} & \textbf{38.81±0.22} & \textbf{57.52±0.13} & \textbf{26.39±0.2}2 & \textbf{27.88±0.26} & \textbf{33.45±0.22} \\
\hline
NRMS   & $\infty$     & 10           & DP-FedRec  & 50.32±0.14 & 22.37±0.11 & 22.18±0.13 & 30.07±0.13 & 50.23±0.06 & 22.79±0.05 & 23.44±0.06 & 28.98±0.05 \\
       & 10           & 10           & DP-FedRec  & 50.18±0.11 & 22.11±0.06 & 21.86±0.10 & 29.74±0.06 & 50.19±0.10 & 22.71±0.06 & 23.34±0.07 & 28.87±0.07 \\
       & 10           & 10           & PrivateRec & \textbf{61.37±0.22} & \textbf{28.77±0.16} & \textbf{30.42±0.21} & \textbf{38.47±0.20} & \textbf{57.00±0.04} & \textbf{26.15±0.14} & \textbf{27.49±0.18} & \textbf{32.74±0.11} \\
\hline
PLM-NR & $\infty$     & 10           & DP-FedRec  & 50.44±0.03 & 21.75±0.06 & 21.40±0.06 & 29.45±0.03 & 50.59±0.01 & 22.61±0.06 & 23.18±0.05 & 28.78±0.06 \\
       & 10           & 10           & DP-FedRec  & 50.16±0.09 & 21.85±0.07 & 21.57±0.09 & 29.48±0.08 & 50.03±0.04 & 22.17±0.06 & 22.70±0.11 & 28.26±0.10 \\
       & 10           & 10           & PrivateRec & \textbf{62.62±0.14} & \textbf{29.74±0.10} & \textbf{31.57±0.11} & \textbf{39.66±0.11} & \textbf{58.69±0.29} & \textbf{26.37±0.11} & \textbf{27.75±0.18} & \textbf{33.57±0.08} \\
\bottomrule[1pt]
\end{tabular}
\end{center}
\end{table*}

Then, we evaluate the privacy-preserving model serving for \textit{PrivateRec} and \textit{DP-FedRec} in Table \ref{tab-serve} under the same serving privacy budget $\epsilon_s$.
First, we can observe that for two versions of baseline, i.e., \textit{DP-FedRec} with or without private training, \textit{PrivateRec} can achieve a significantly better performance across all models and datasets.
\begin{figure*}[t]
    \centering
    \includegraphics[width=\textwidth]{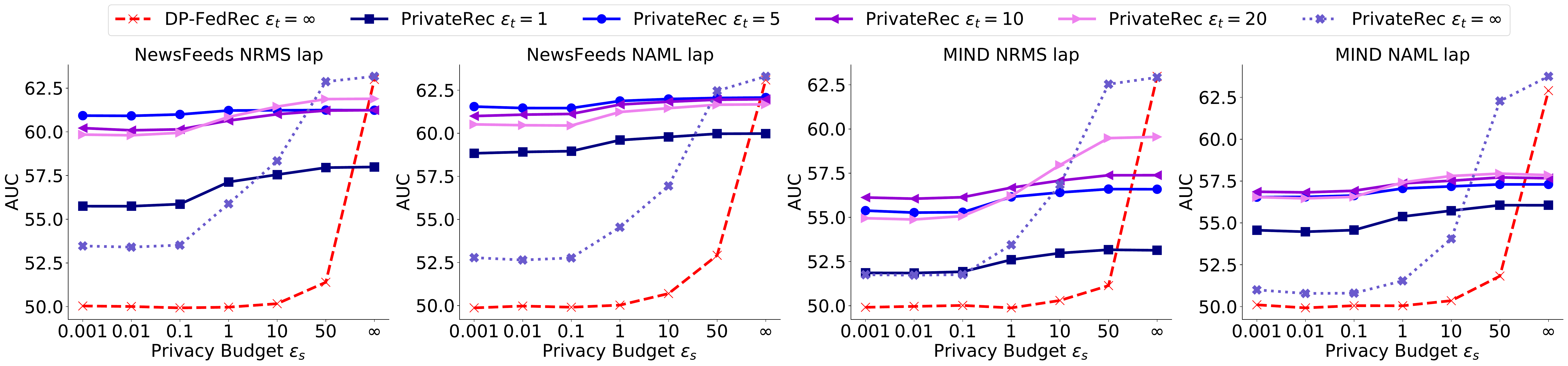}
    \caption{Performance of privacy-preserving model serving. ($\epsilon_t=\infty$ for \textit{DP-FedRec} and $\epsilon_t=10, B=5$ for \textit{PrivateRec})}
    \label{fig-serve}
\end{figure*}

\begin{figure*}[t]
\centering
\includegraphics[trim=5 10 5 18, clip, width=\textwidth]{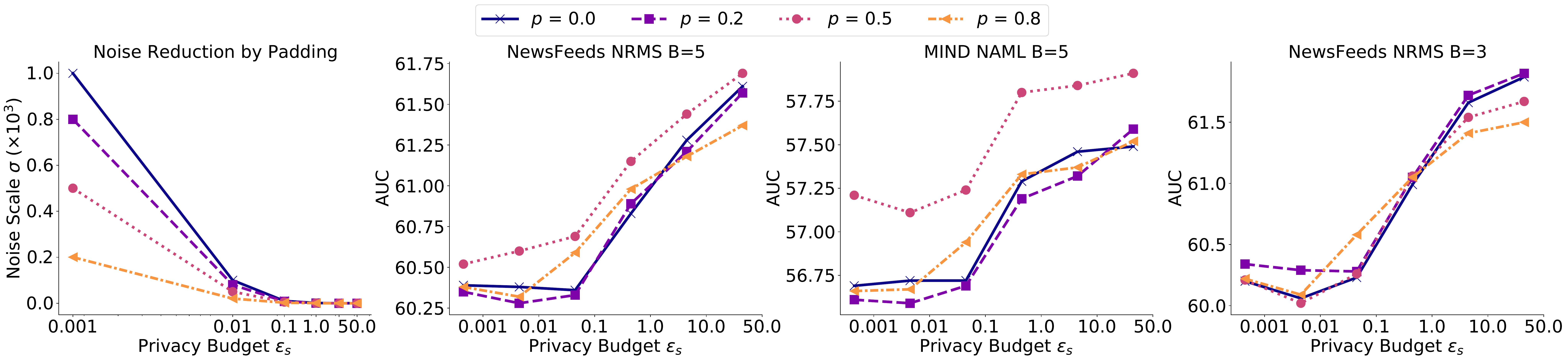}
\caption{Effect of privacy amplification by behavior padding.}
\label{fig-serve-pad}
\end{figure*}

Then we investigate the privacy-utility trade-off in Fig.~\ref{fig-serve}.
First, we can observe that the model serving performance of \textit{PrivateRec} is consistently better than \textit{DP-FedRec} under various serving privacy budgets $\epsilon_s$.
Second, we find the a proper amount of noise injected in \textit{PrivateFec} training can help the privacy-preserving model serving.
From Fig.~\ref{fig-serve}, the performance of \textit{PrivateRec} with $\epsilon=\infty$ in the dashed line cannot outperform privacy-preserving training \textit{PrivateRec} for $\epsilon_s$ from 0.001 to 10.
This fact emphasizes our claim that the design of privacy-preserving federated training should co-design with the privacy-preserving model serving as an end-to-end solution for the federated recommendation system.
If $\epsilon_t$ is too small (e.g, 1), the performance upper bound under $(\epsilon_s=\infty)$ is limited.
If $\epsilon_t$ is too large (e.g., 20), the performance upper bound is higher.
And under a smaller $\epsilon_s$, it is not as robust as the performance with a moderate $\epsilon_t=10$.
Thus, we conclude that the $\epsilon_t$ can be the knob to tune the privacy-utility trade-off in privacy-preserving model serving.
More importantly, we find that the utility gain of \textit{PrivateRec} in federated training can maintain for privacy-preserving model serving, whereas \textit{DP-FedRec} cannot.
As shown in Fig.~\ref{fig-serve}, the utility of \textit{FedRec} with $\epsilon_s=\infty$ is better than \textit{PrivateRec}.
However, the model serving utility is catastrophically ruined by the DP noise.

Next, we evaluate the effect of privacy amplification by behavior padding in Fig.~\ref{fig-serve-pad}.
The motivation for padding the sequence of users' historical behavior representations is to reduce the amount of noise with the privacy amplification effect by nullification.
Thus, in the first sub-figure, we show the required noise perturbation scales for the same serving privacy budget $\epsilon_s$ w.r.t different padding ratios.
We can see that the required noise is reduced with a larger padding ratio.
Additionally, this effect is more significant when the privacy requirement for serving is stricter (i.e, with a smaller $\epsilon_s$).
Generally, $p$ controls the trade-off between the personalized interest and the generalized interest as well as the strength of noise reduction.
From the right three sub-figures, we can observe that if $p$ is too small, the noise reduction by the privacy amplification is negligible.
If $p$ is too large, the personalized information is erased, which reduces the utility of model serving.
Concretely, the best $p$ for \textit{PrivateRec} is 0.5 when $B=5$ and 0.2 when $B=3$.
This is reasonable because a smaller $B$ indicates more coarse-grained interest summarization, and the information for each basic vector is more abstract.
Thus, for \textit{PrivateRec} with a smaller $B$, the personalized interest information is more important, thus a smaller $p$ is preferred.

\begin{figure}[h]
\centering
\includegraphics[width=0.5\textwidth]{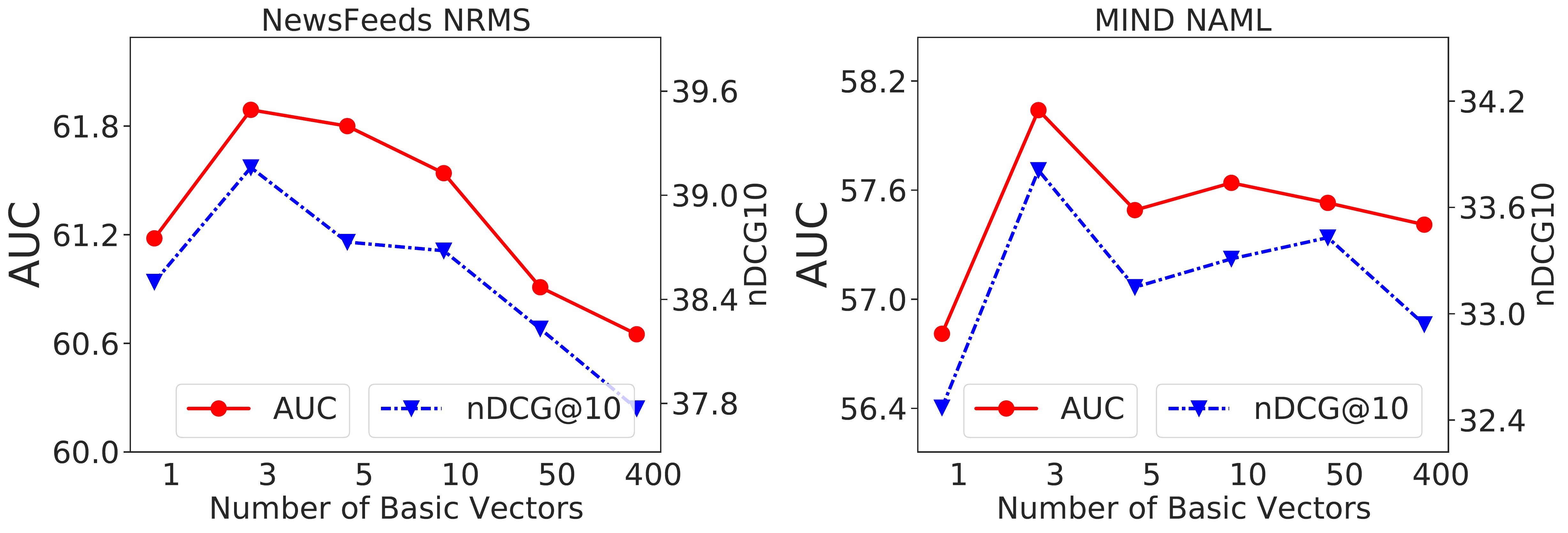}
\caption{Influence of the number of basic vectors on the training performance of \textit{PrivateRec} with $\epsilon_t=10, \epsilon_s=\infty$.}
\label{fig-train-inf-nb}
\end{figure}
Last, we evaluate the influence of the number of basic vectors in Fig.~\ref{fig-train-inf-nb} for \textit{PrivateRec} federated training.
We observe that the best number of basic vectors for \textit{NewsFeeds} and \textit{MIND} datasets are $B=3$.
If $B$ is too small, the basic vectors might be too coarse-grained and cannot express the personal user interest.
If $B$ is too large, the dimension of an attention vector $\mathbf{\alpha}$ is large, which results in an increased amount of noise.

\section{Conclusions}
In this paper, we are the first to rethink and formulate the privacy definitions in federated news recommendation.
Then, we propose a differentially private federated news recommendation framework \textit{PrivateRec}, which can achieve a better utility for training and serving under a formal privacy definition.
To avoid the dimension-dependent noise and improve utility in privacy-preserving federated training and serving, we
decompose the high-dimensional and privacy-sensitive user embedding into a combination of public basic vectors and add noise to the combination coefficients.
We further protect privacy in training with a label perturbation module.
In addition, to further reduce noise, we utilize the amplification effect by randomly padding user historical behavior representations.
Experiments on two real-world news recommendation datasets validate our method's effectiveness and utility improvement on the model training and serving stage.
However, \textit{PrivateRec} are designed for recommendation models with individual user encoder and item encoder.
Thus, we plan to extend the formulated privacy definition and privacy mechanisms for general recommendation models.

\bibliographystyle{ACM-Reference-Format}
\bibliography{reference}


\begin{thebibliography}{50}


\ifx \showCODEN    \undefined \def \showCODEN     #1{\unskip}     \fi
\ifx \showDOI      \undefined \def \showDOI       #1{#1}\fi
\ifx \showISBNx    \undefined \def \showISBNx     #1{\unskip}     \fi
\ifx \showISBNxiii \undefined \def \showISBNxiii  #1{\unskip}     \fi
\ifx \showISSN     \undefined \def \showISSN      #1{\unskip}     \fi
\ifx \showLCCN     \undefined \def \showLCCN      #1{\unskip}     \fi
\ifx \shownote     \undefined \def \shownote      #1{#1}          \fi
\ifx \showarticletitle \undefined \def \showarticletitle #1{#1}   \fi
\ifx \showURL      \undefined \def \showURL       {\relax}        \fi
\providecommand\bibfield[2]{#2}
\providecommand\bibinfo[2]{#2}
\providecommand\natexlab[1]{#1}
\providecommand\showeprint[2][]{arXiv:#2}

\bibitem[\protect\citeauthoryear{Abadi, Chu, Goodfellow, McMahan, Mironov,
  Talwar, and Zhang}{Abadi et~al\mbox{.}}{2016}]%
        {abadi2016deep}
\bibfield{author}{\bibinfo{person}{Martin Abadi}, \bibinfo{person}{Andy Chu},
  \bibinfo{person}{Ian Goodfellow}, \bibinfo{person}{H~Brendan McMahan},
  \bibinfo{person}{Ilya Mironov}, \bibinfo{person}{Kunal Talwar}, {and}
  \bibinfo{person}{Li Zhang}.} \bibinfo{year}{2016}\natexlab{}.
\newblock \showarticletitle{Deep learning with differential privacy}. In
  \bibinfo{booktitle}{\emph{CCS}}. \bibinfo{pages}{308--318}.
\newblock


\bibitem[\protect\citeauthoryear{Agarap}{Agarap}{2018}]%
        {agarap2018deep}
\bibfield{author}{\bibinfo{person}{Abien~Fred Agarap}.}
  \bibinfo{year}{2018}\natexlab{}.
\newblock \showarticletitle{Deep learning using rectified linear units (relu)}.
\newblock \bibinfo{journal}{\emph{arXiv preprint arXiv:1803.08375}}
  (\bibinfo{year}{2018}).
\newblock


\bibitem[\protect\citeauthoryear{Ammad-Ud-Din, Ivannikova, Khan, Oyomno, Fu,
  Tan, and Flanagan}{Ammad-Ud-Din et~al\mbox{.}}{2019}]%
        {ammad2019federated}
\bibfield{author}{\bibinfo{person}{Muhammad Ammad-Ud-Din},
  \bibinfo{person}{Elena Ivannikova}, \bibinfo{person}{Suleiman~A Khan},
  \bibinfo{person}{Were Oyomno}, \bibinfo{person}{Qiang Fu},
  \bibinfo{person}{Kuan~Eeik Tan}, {and} \bibinfo{person}{Adrian Flanagan}.}
  \bibinfo{year}{2019}\natexlab{}.
\newblock \showarticletitle{Federated collaborative filtering for
  privacy-preserving personalized recommendation system}.
\newblock \bibinfo{journal}{\emph{arXiv preprint arXiv:1901.09888}}
  (\bibinfo{year}{2019}).
\newblock


\bibitem[\protect\citeauthoryear{An, Wu, Wang, Di, Huang, and Xie}{An
  et~al\mbox{.}}{2019a}]%
        {an2019neural}
\bibfield{author}{\bibinfo{person}{Mingxiao An}, \bibinfo{person}{Fangzhao Wu},
  \bibinfo{person}{Heyuan Wang}, \bibinfo{person}{Tao Di},
  \bibinfo{person}{Jianqiang Huang}, {and} \bibinfo{person}{Xing Xie}.}
  \bibinfo{year}{2019}\natexlab{a}.
\newblock \showarticletitle{Neural ctr prediction for native ad}. In
  \bibinfo{booktitle}{\emph{CCL}}. Springer, \bibinfo{pages}{600--612}.
\newblock


\bibitem[\protect\citeauthoryear{An, Wu, Wu, Zhang, Liu, and Xie}{An
  et~al\mbox{.}}{2019b}]%
        {an2019neural2}
\bibfield{author}{\bibinfo{person}{Mingxiao An}, \bibinfo{person}{Fangzhao Wu},
  \bibinfo{person}{Chuhan Wu}, \bibinfo{person}{Kun Zhang},
  \bibinfo{person}{Zheng Liu}, {and} \bibinfo{person}{Xing Xie}.}
  \bibinfo{year}{2019}\natexlab{b}.
\newblock \showarticletitle{Neural news recommendation with long-and short-term
  user representations}. In \bibinfo{booktitle}{\emph{ACL}}.
  \bibinfo{pages}{336--345}.
\newblock


\bibitem[\protect\citeauthoryear{Balle and Wang}{Balle and Wang}{2018}]%
        {balle2018improving}
\bibfield{author}{\bibinfo{person}{Borja Balle} {and} \bibinfo{person}{Yu-Xiang
  Wang}.} \bibinfo{year}{2018}\natexlab{}.
\newblock \showarticletitle{Improving the gaussian mechanism for differential
  privacy: Analytical calibration and optimal denoising}. In
  \bibinfo{booktitle}{\emph{International Conference on Machine Learning}}.
  PMLR, \bibinfo{pages}{394--403}.
\newblock


\bibitem[\protect\citeauthoryear{Bonawitz, Ivanov, Kreuter, Marcedone, McMahan,
  Patel, Ramage, Segal, and Seth}{Bonawitz et~al\mbox{.}}{2017}]%
        {bonawitz2017practical}
\bibfield{author}{\bibinfo{person}{Keith Bonawitz}, \bibinfo{person}{Vladimir
  Ivanov}, \bibinfo{person}{Ben Kreuter}, \bibinfo{person}{Antonio Marcedone},
  \bibinfo{person}{H~Brendan McMahan}, \bibinfo{person}{Sarvar Patel},
  \bibinfo{person}{Daniel Ramage}, \bibinfo{person}{Aaron Segal}, {and}
  \bibinfo{person}{Karn Seth}.} \bibinfo{year}{2017}\natexlab{}.
\newblock \showarticletitle{Practical secure aggregation for privacy-preserving
  machine learning}. In \bibinfo{booktitle}{\emph{CCS}}.
  \bibinfo{pages}{1175--1191}.
\newblock


\bibitem[\protect\citeauthoryear{Chai, Wang, Chen, and Yang}{Chai
  et~al\mbox{.}}{2020}]%
        {chai2020secure}
\bibfield{author}{\bibinfo{person}{Di Chai}, \bibinfo{person}{Leye Wang},
  \bibinfo{person}{Kai Chen}, {and} \bibinfo{person}{Qiang Yang}.}
  \bibinfo{year}{2020}\natexlab{}.
\newblock \showarticletitle{Secure federated matrix factorization}.
\newblock \bibinfo{journal}{\emph{IEEE Intelligent Systems}}
  (\bibinfo{year}{2020}).
\newblock


\bibitem[\protect\citeauthoryear{Chen, Zhang, Tung, Kankanhalli, and Chen}{Chen
  et~al\mbox{.}}{2020a}]%
        {chen2020robust}
\bibfield{author}{\bibinfo{person}{Chen Chen}, \bibinfo{person}{Jingfeng
  Zhang}, \bibinfo{person}{Anthony~KH Tung}, \bibinfo{person}{Mohan
  Kankanhalli}, {and} \bibinfo{person}{Gang Chen}.}
  \bibinfo{year}{2020}\natexlab{a}.
\newblock \showarticletitle{Robust federated recommendation system}.
\newblock \bibinfo{journal}{\emph{arXiv preprint arXiv:2006.08259}}
  (\bibinfo{year}{2020}).
\newblock


\bibitem[\protect\citeauthoryear{Chen, Zhou, Wu, Fang, Wang, Qi, and
  Zheng}{Chen et~al\mbox{.}}{2020b}]%
        {chen2020practical}
\bibfield{author}{\bibinfo{person}{Chaochao Chen}, \bibinfo{person}{Jun Zhou},
  \bibinfo{person}{Bingzhe Wu}, \bibinfo{person}{Wenjing Fang},
  \bibinfo{person}{Li Wang}, \bibinfo{person}{Yuan Qi}, {and}
  \bibinfo{person}{Xiaolin Zheng}.} \bibinfo{year}{2020}\natexlab{b}.
\newblock \showarticletitle{Practical privacy preserving POI recommendation}.
\newblock \bibinfo{journal}{\emph{TIST}} \bibinfo{volume}{11},
  \bibinfo{number}{5} (\bibinfo{year}{2020}), \bibinfo{pages}{1--20}.
\newblock


\bibitem[\protect\citeauthoryear{Covington, Adams, and Sargin}{Covington
  et~al\mbox{.}}{2016}]%
        {covington2016deep}
\bibfield{author}{\bibinfo{person}{Paul Covington}, \bibinfo{person}{Jay
  Adams}, {and} \bibinfo{person}{Emre Sargin}.}
  \bibinfo{year}{2016}\natexlab{}.
\newblock \showarticletitle{Deep neural networks for youtube recommendations}.
  In \bibinfo{booktitle}{\emph{RecSys}}. \bibinfo{pages}{191--198}.
\newblock


\bibitem[\protect\citeauthoryear{Dwork, McSherry, Nissim, and Smith}{Dwork
  et~al\mbox{.}}{2006}]%
        {dwork2006calibrating}
\bibfield{author}{\bibinfo{person}{Cynthia Dwork}, \bibinfo{person}{Frank
  McSherry}, \bibinfo{person}{Kobbi Nissim}, {and} \bibinfo{person}{Adam
  Smith}.} \bibinfo{year}{2006}\natexlab{}.
\newblock \showarticletitle{Calibrating noise to sensitivity in private data
  analysis}. In \bibinfo{booktitle}{\emph{TCC}}. Springer,
  \bibinfo{pages}{265--284}.
\newblock


\bibitem[\protect\citeauthoryear{Dwork, Roth, et~al\mbox{.}}{Dwork
  et~al\mbox{.}}{2014}]%
        {dwork2014algorithmic}
\bibfield{author}{\bibinfo{person}{Cynthia Dwork}, \bibinfo{person}{Aaron
  Roth}, {et~al\mbox{.}}} \bibinfo{year}{2014}\natexlab{}.
\newblock \showarticletitle{The algorithmic foundations of differential
  privacy.}
\newblock \bibinfo{journal}{\emph{Found. Trends Theor. Comput. Sci.}}
  \bibinfo{volume}{9}, \bibinfo{number}{3-4} (\bibinfo{year}{2014}),
  \bibinfo{pages}{211--407}.
\newblock


\bibitem[\protect\citeauthoryear{Gao, Tan, Ju, Zheng, and Yang}{Gao
  et~al\mbox{.}}{2020}]%
        {gao2020privacy}
\bibfield{author}{\bibinfo{person}{Dashan Gao}, \bibinfo{person}{Ben Tan},
  \bibinfo{person}{Ce Ju}, \bibinfo{person}{Vincent~W Zheng}, {and}
  \bibinfo{person}{Qiang Yang}.} \bibinfo{year}{2020}\natexlab{}.
\newblock \showarticletitle{Privacy threats against federated matrix
  factorization}.
\newblock \bibinfo{journal}{\emph{arXiv preprint arXiv:2007.01587}}
  (\bibinfo{year}{2020}).
\newblock


\bibitem[\protect\citeauthoryear{Glorot, Bordes, and Bengio}{Glorot
  et~al\mbox{.}}{2011}]%
        {glorot2011deep}
\bibfield{author}{\bibinfo{person}{Xavier Glorot}, \bibinfo{person}{Antoine
  Bordes}, {and} \bibinfo{person}{Yoshua Bengio}.}
  \bibinfo{year}{2011}\natexlab{}.
\newblock \showarticletitle{Deep sparse rectifier neural networks}. JMLR
  Workshop and Conference Proceedings, \bibinfo{pages}{315--323}.
\newblock


\bibitem[\protect\citeauthoryear{He, Du, Wang, Tian, Tang, and Chua}{He
  et~al\mbox{.}}{2018}]%
        {he2018outer}
\bibfield{author}{\bibinfo{person}{Xiangnan He}, \bibinfo{person}{Xiaoyu Du},
  \bibinfo{person}{Xiang Wang}, \bibinfo{person}{Feng Tian},
  \bibinfo{person}{Jinhui Tang}, {and} \bibinfo{person}{Tat-Seng Chua}.}
  \bibinfo{year}{2018}\natexlab{}.
\newblock \showarticletitle{Outer product-based neural collaborative
  filtering}.
\newblock \bibinfo{journal}{\emph{arXiv preprint arXiv:1808.03912}}
  (\bibinfo{year}{2018}).
\newblock


\bibitem[\protect\citeauthoryear{Huang, Li, Bai, Wang, Bai, and Wang}{Huang
  et~al\mbox{.}}{2020}]%
        {huang2020federated}
\bibfield{author}{\bibinfo{person}{Mingkai Huang}, \bibinfo{person}{Hao Li},
  \bibinfo{person}{Bing Bai}, \bibinfo{person}{Chang Wang},
  \bibinfo{person}{Kun Bai}, {and} \bibinfo{person}{Fei Wang}.}
  \bibinfo{year}{2020}\natexlab{}.
\newblock \showarticletitle{A Federated Multi-View Deep Learning Framework for
  Privacy-Preserving Recommendations}.
\newblock \bibinfo{journal}{\emph{arXiv preprint arXiv:2008.10808}}
  (\bibinfo{year}{2020}).
\newblock


\bibitem[\protect\citeauthoryear{Li, Liu, Wu, Xu, Zhao, Huang, Kang, Chen, Li,
  and Lee}{Li et~al\mbox{.}}{2019}]%
        {li2019multi}
\bibfield{author}{\bibinfo{person}{Chao Li}, \bibinfo{person}{Zhiyuan Liu},
  \bibinfo{person}{Mengmeng Wu}, \bibinfo{person}{Yuchi Xu},
  \bibinfo{person}{Huan Zhao}, \bibinfo{person}{Pipei Huang},
  \bibinfo{person}{Guoliang Kang}, \bibinfo{person}{Qiwei Chen},
  \bibinfo{person}{Wei Li}, {and} \bibinfo{person}{Dik~Lun Lee}.}
  \bibinfo{year}{2019}\natexlab{}.
\newblock \showarticletitle{Multi-interest network with dynamic routing for
  recommendation at Tmall}. In \bibinfo{booktitle}{\emph{CIKM}}.
  \bibinfo{pages}{2615--2623}.
\newblock


\bibitem[\protect\citeauthoryear{Li, Lyu, Su, and Yang}{Li
  et~al\mbox{.}}{2016}]%
        {li2016differential}
\bibfield{author}{\bibinfo{person}{Ninghui Li}, \bibinfo{person}{Min Lyu},
  \bibinfo{person}{Dong Su}, {and} \bibinfo{person}{Weining Yang}.}
  \bibinfo{year}{2016}\natexlab{}.
\newblock \showarticletitle{Differential privacy: From theory to practice}.
\newblock \bibinfo{journal}{\emph{Synthesis Lectures on Information Security,
  Privacy, \& Trust}} \bibinfo{volume}{8}, \bibinfo{number}{4}
  (\bibinfo{year}{2016}), \bibinfo{pages}{1--138}.
\newblock


\bibitem[\protect\citeauthoryear{Li, Song, and Fragouli}{Li
  et~al\mbox{.}}{2020}]%
        {li2020federated}
\bibfield{author}{\bibinfo{person}{Tan Li}, \bibinfo{person}{Linqi Song}, {and}
  \bibinfo{person}{Christina Fragouli}.} \bibinfo{year}{2020}\natexlab{}.
\newblock \showarticletitle{Federated recommendation system via differential
  privacy}. In \bibinfo{booktitle}{\emph{ISIT}}. IEEE,
  \bibinfo{pages}{2592--2597}.
\newblock


\bibitem[\protect\citeauthoryear{Li, Ding, Zhang, Li, and Zhou}{Li
  et~al\mbox{.}}{2021}]%
        {li2021federatedMF}
\bibfield{author}{\bibinfo{person}{Zitao Li}, \bibinfo{person}{Bolin Ding},
  \bibinfo{person}{Ce Zhang}, \bibinfo{person}{Ninghui Li}, {and}
  \bibinfo{person}{Jingren Zhou}.} \bibinfo{year}{2021}\natexlab{}.
\newblock \showarticletitle{Federated matrix factorization with privacy
  guarantee}.
\newblock \bibinfo{journal}{\emph{Proceedings of the VLDB Endowment}}
  \bibinfo{volume}{15}, \bibinfo{number}{4} (\bibinfo{year}{2021}),
  \bibinfo{pages}{900--913}.
\newblock


\bibitem[\protect\citeauthoryear{Liang, Pan, and Ming}{Liang
  et~al\mbox{.}}{2021}]%
        {liang2021fedrec++}
\bibfield{author}{\bibinfo{person}{Feng Liang}, \bibinfo{person}{Weike Pan},
  {and} \bibinfo{person}{Zhong Ming}.} \bibinfo{year}{2021}\natexlab{}.
\newblock \showarticletitle{Fedrec++: Lossless federated recommendation with
  explicit feedback}. In \bibinfo{booktitle}{\emph{Proceedings of the AAAI
  conference on artificial intelligence}}, Vol.~\bibinfo{volume}{35}.
  \bibinfo{pages}{4224--4231}.
\newblock


\bibitem[\protect\citeauthoryear{Lin, Liang, Pan, and Ming}{Lin
  et~al\mbox{.}}{2020}]%
        {lin2020fedrec}
\bibfield{author}{\bibinfo{person}{Guanyu Lin}, \bibinfo{person}{Feng Liang},
  \bibinfo{person}{Weike Pan}, {and} \bibinfo{person}{Zhong Ming}.}
  \bibinfo{year}{2020}\natexlab{}.
\newblock \showarticletitle{Fedrec: Federated recommendation with explicit
  feedback}.
\newblock \bibinfo{journal}{\emph{IEEE Intelligent Systems}}
  (\bibinfo{year}{2020}).
\newblock


\bibitem[\protect\citeauthoryear{Lyu, He, and Li}{Lyu et~al\mbox{.}}{2020}]%
        {lyu2020differentially}
\bibfield{author}{\bibinfo{person}{Lingjuan Lyu}, \bibinfo{person}{Xuanli He},
  {and} \bibinfo{person}{Yitong Li}.} \bibinfo{year}{2020}\natexlab{}.
\newblock \showarticletitle{Differentially private representation for nlp:
  Formal guarantee and an empirical study on privacy and fairness}.
\newblock \bibinfo{journal}{\emph{arXiv preprint arXiv:2010.01285}}
  (\bibinfo{year}{2020}).
\newblock


\bibitem[\protect\citeauthoryear{McMahan, Moore, Ramage, Hampson, and
  y~Arcas}{McMahan et~al\mbox{.}}{2017a}]%
        {mcmahan2017communication}
\bibfield{author}{\bibinfo{person}{Brendan McMahan}, \bibinfo{person}{Eider
  Moore}, \bibinfo{person}{Daniel Ramage}, \bibinfo{person}{Seth Hampson},
  {and} \bibinfo{person}{Blaise~Aguera y Arcas}.}
  \bibinfo{year}{2017}\natexlab{a}.
\newblock \showarticletitle{Communication-efficient learning of deep networks
  from decentralized data}. In \bibinfo{booktitle}{\emph{Artificial
  intelligence and statistics}}. PMLR, \bibinfo{pages}{1273--1282}.
\newblock


\bibitem[\protect\citeauthoryear{McMahan, Ramage, Talwar, and Zhang}{McMahan
  et~al\mbox{.}}{2017b}]%
        {mcmahan2017learning}
\bibfield{author}{\bibinfo{person}{H~Brendan McMahan}, \bibinfo{person}{Daniel
  Ramage}, \bibinfo{person}{Kunal Talwar}, {and} \bibinfo{person}{Li Zhang}.}
  \bibinfo{year}{2017}\natexlab{b}.
\newblock \showarticletitle{Learning differentially private recurrent language
  models}.
\newblock \bibinfo{journal}{\emph{arXiv preprint arXiv:1710.06963}}
  (\bibinfo{year}{2017}).
\newblock


\bibitem[\protect\citeauthoryear{McSherry and Talwar}{McSherry and
  Talwar}{2007}]%
        {mcsherry2007mechanism}
\bibfield{author}{\bibinfo{person}{Frank McSherry} {and} \bibinfo{person}{Kunal
  Talwar}.} \bibinfo{year}{2007}\natexlab{}.
\newblock \showarticletitle{Mechanism design via differential privacy}. In
  \bibinfo{booktitle}{\emph{FOCS}}. IEEE, \bibinfo{pages}{94--103}.
\newblock


\bibitem[\protect\citeauthoryear{Muhammad, Wang, O'Reilly-Morgan, Tragos,
  Smyth, Hurley, Geraci, and Lawlor}{Muhammad et~al\mbox{.}}{2020}]%
        {muhammad2020fedfast}
\bibfield{author}{\bibinfo{person}{Khalil Muhammad}, \bibinfo{person}{Qinqin
  Wang}, \bibinfo{person}{Diarmuid O'Reilly-Morgan}, \bibinfo{person}{Elias
  Tragos}, \bibinfo{person}{Barry Smyth}, \bibinfo{person}{Neil Hurley},
  \bibinfo{person}{James Geraci}, {and} \bibinfo{person}{Aonghus Lawlor}.}
  \bibinfo{year}{2020}\natexlab{}.
\newblock \showarticletitle{Fedfast: Going beyond average for faster training
  of federated recommender systems}. In \bibinfo{booktitle}{\emph{SIGKDD}}.
  \bibinfo{pages}{1234--1242}.
\newblock


\bibitem[\protect\citeauthoryear{Nasr, Shokri, and Houmansadr}{Nasr
  et~al\mbox{.}}{2019}]%
        {nasr2019comprehensive}
\bibfield{author}{\bibinfo{person}{Milad Nasr}, \bibinfo{person}{Reza Shokri},
  {and} \bibinfo{person}{Amir Houmansadr}.} \bibinfo{year}{2019}\natexlab{}.
\newblock \showarticletitle{Comprehensive privacy analysis of deep learning:
  Passive and active white-box inference attacks against centralized and
  federated learning}. In \bibinfo{booktitle}{\emph{SP}}. IEEE,
  \bibinfo{pages}{739--753}.
\newblock


\bibitem[\protect\citeauthoryear{Nguy{\^e}n, Xiao, Yang, Hui, Shin, and
  Shin}{Nguy{\^e}n et~al\mbox{.}}{2016}]%
        {nguyen2016collecting}
\bibfield{author}{\bibinfo{person}{Th{\^o}ng~T Nguy{\^e}n},
  \bibinfo{person}{Xiaokui Xiao}, \bibinfo{person}{Yin Yang},
  \bibinfo{person}{Siu~Cheung Hui}, \bibinfo{person}{Hyejin Shin}, {and}
  \bibinfo{person}{Junbum Shin}.} \bibinfo{year}{2016}\natexlab{}.
\newblock \showarticletitle{Collecting and analyzing data from smart device
  users with local differential privacy}.
\newblock \bibinfo{journal}{\emph{arXiv preprint arXiv:1606.05053}}
  (\bibinfo{year}{2016}).
\newblock


\bibitem[\protect\citeauthoryear{Okura, Tagami, Ono, and Tajima}{Okura
  et~al\mbox{.}}{2017}]%
        {okura2017embedding}
\bibfield{author}{\bibinfo{person}{Shumpei Okura}, \bibinfo{person}{Yukihiro
  Tagami}, \bibinfo{person}{Shingo Ono}, {and} \bibinfo{person}{Akira Tajima}.}
  \bibinfo{year}{2017}\natexlab{}.
\newblock \showarticletitle{Embedding-based news recommendation for millions of
  users}. In \bibinfo{booktitle}{\emph{SIGKDD}}. \bibinfo{pages}{1933--1942}.
\newblock


\bibitem[\protect\citeauthoryear{Qi, Wu, Wu, Huang, and Xie}{Qi
  et~al\mbox{.}}{2020}]%
        {qi2020privacy}
\bibfield{author}{\bibinfo{person}{Tao Qi}, \bibinfo{person}{Fangzhao Wu},
  \bibinfo{person}{Chuhan Wu}, \bibinfo{person}{Yongfeng Huang}, {and}
  \bibinfo{person}{Xing Xie}.} \bibinfo{year}{2020}\natexlab{}.
\newblock \showarticletitle{Privacy-Preserving News Recommendation Model
  Training via Federated Learning}.
\newblock  (\bibinfo{year}{2020}).
\newblock


\bibitem[\protect\citeauthoryear{Qi, Wu, Wu, Huang, and Xie}{Qi
  et~al\mbox{.}}{2021}]%
        {qi2021uni}
\bibfield{author}{\bibinfo{person}{Tao Qi}, \bibinfo{person}{Fangzhao Wu},
  \bibinfo{person}{Chuhan Wu}, \bibinfo{person}{Yongfeng Huang}, {and}
  \bibinfo{person}{Xing Xie}.} \bibinfo{year}{2021}\natexlab{}.
\newblock \showarticletitle{Uni-FedRec: A Unified Privacy-Preserving News
  Recommendation Framework for Model Training and Online Serving}.
\newblock \bibinfo{journal}{\emph{arXiv preprint arXiv:2109.05236}}
  (\bibinfo{year}{2021}).
\newblock


\bibitem[\protect\citeauthoryear{Reddi, Charles, Zaheer, Garrett, Rush,
  Kone{\v{c}}n{\`y}, Kumar, and McMahan}{Reddi et~al\mbox{.}}{2020}]%
        {reddi2020adaptive}
\bibfield{author}{\bibinfo{person}{Sashank Reddi}, \bibinfo{person}{Zachary
  Charles}, \bibinfo{person}{Manzil Zaheer}, \bibinfo{person}{Zachary Garrett},
  \bibinfo{person}{Keith Rush}, \bibinfo{person}{Jakub Kone{\v{c}}n{\`y}},
  \bibinfo{person}{Sanjiv Kumar}, {and} \bibinfo{person}{H~Brendan McMahan}.}
  \bibinfo{year}{2020}\natexlab{}.
\newblock \showarticletitle{Adaptive federated optimization}.
\newblock \bibinfo{journal}{\emph{arXiv preprint arXiv:2003.00295}}
  (\bibinfo{year}{2020}).
\newblock


\bibitem[\protect\citeauthoryear{Rendle}{Rendle}{2010}]%
        {rendle2010factorization}
\bibfield{author}{\bibinfo{person}{Steffen Rendle}.}
  \bibinfo{year}{2010}\natexlab{}.
\newblock \showarticletitle{Factorization machines}. In
  \bibinfo{booktitle}{\emph{ICDM}}. IEEE, \bibinfo{pages}{995--1000}.
\newblock


\bibitem[\protect\citeauthoryear{Shin, Kim, Shin, and Xiao}{Shin
  et~al\mbox{.}}{2018}]%
        {shin2018privacy}
\bibfield{author}{\bibinfo{person}{Hyejin Shin}, \bibinfo{person}{Sungwook
  Kim}, \bibinfo{person}{Junbum Shin}, {and} \bibinfo{person}{Xiaokui Xiao}.}
  \bibinfo{year}{2018}\natexlab{}.
\newblock \showarticletitle{Privacy enhanced matrix factorization for
  recommendation with local differential privacy}.
\newblock \bibinfo{journal}{\emph{TKDE}} \bibinfo{volume}{30},
  \bibinfo{number}{9} (\bibinfo{year}{2018}), \bibinfo{pages}{1770--1782}.
\newblock


\bibitem[\protect\citeauthoryear{Shokri and Shmatikov}{Shokri and
  Shmatikov}{2015}]%
        {shokri2015privacy}
\bibfield{author}{\bibinfo{person}{Reza Shokri} {and} \bibinfo{person}{Vitaly
  Shmatikov}.} \bibinfo{year}{2015}\natexlab{}.
\newblock \showarticletitle{Privacy-preserving deep learning}. In
  \bibinfo{booktitle}{\emph{Proceedings of the 22nd ACM SIGSAC conference on
  computer and communications security}}. \bibinfo{pages}{1310--1321}.
\newblock


\bibitem[\protect\citeauthoryear{Truex, Liu, Chow, Gursoy, and Wei}{Truex
  et~al\mbox{.}}{2020}]%
        {truex2020ldp}
\bibfield{author}{\bibinfo{person}{Stacey Truex}, \bibinfo{person}{Ling Liu},
  \bibinfo{person}{Ka-Ho Chow}, \bibinfo{person}{Mehmet~Emre Gursoy}, {and}
  \bibinfo{person}{Wenqi Wei}.} \bibinfo{year}{2020}\natexlab{}.
\newblock \showarticletitle{LDP-Fed: Federated learning with local differential
  privacy}. In \bibinfo{booktitle}{\emph{Proceedings of the Third ACM
  International Workshop on Edge Systems, Analytics and Networking}}.
  \bibinfo{pages}{61--66}.
\newblock


\bibitem[\protect\citeauthoryear{Wang, Zhang, Xie, and Guo}{Wang
  et~al\mbox{.}}{2018b}]%
        {wang2018dkn}
\bibfield{author}{\bibinfo{person}{Hongwei Wang}, \bibinfo{person}{Fuzheng
  Zhang}, \bibinfo{person}{Xing Xie}, {and} \bibinfo{person}{Minyi Guo}.}
  \bibinfo{year}{2018}\natexlab{b}.
\newblock \showarticletitle{DKN: Deep knowledge-aware network for news
  recommendation}. In \bibinfo{booktitle}{\emph{WWW}}.
  \bibinfo{pages}{1835--1844}.
\newblock


\bibitem[\protect\citeauthoryear{Wang, Zhang, Bao, Zhu, Cao, and Yu}{Wang
  et~al\mbox{.}}{2018a}]%
        {wang2018not}
\bibfield{author}{\bibinfo{person}{Ji Wang}, \bibinfo{person}{Jianguo Zhang},
  \bibinfo{person}{Weidong Bao}, \bibinfo{person}{Xiaomin Zhu},
  \bibinfo{person}{Bokai Cao}, {and} \bibinfo{person}{Philip~S Yu}.}
  \bibinfo{year}{2018}\natexlab{a}.
\newblock \showarticletitle{Not just privacy: Improving performance of private
  deep learning in mobile cloud}. In \bibinfo{booktitle}{\emph{SIGKDD}}.
  \bibinfo{pages}{2407--2416}.
\newblock


\bibitem[\protect\citeauthoryear{Wu, Wu, An, Huang, Huang, and Xie}{Wu
  et~al\mbox{.}}{2019a}]%
        {wu2019neural}
\bibfield{author}{\bibinfo{person}{Chuhan Wu}, \bibinfo{person}{Fangzhao Wu},
  \bibinfo{person}{Mingxiao An}, \bibinfo{person}{Jianqiang Huang},
  \bibinfo{person}{Yongfeng Huang}, {and} \bibinfo{person}{Xing Xie}.}
  \bibinfo{year}{2019}\natexlab{a}.
\newblock \showarticletitle{Neural news recommendation with attentive
  multi-view learning}.
\newblock \bibinfo{journal}{\emph{arXiv preprint arXiv:1907.05576}}
  (\bibinfo{year}{2019}).
\newblock


\bibitem[\protect\citeauthoryear{Wu, Wu, An, Huang, Huang, and Xie}{Wu
  et~al\mbox{.}}{2019b}]%
        {wu2019npa}
\bibfield{author}{\bibinfo{person}{Chuhan Wu}, \bibinfo{person}{Fangzhao Wu},
  \bibinfo{person}{Mingxiao An}, \bibinfo{person}{Jianqiang Huang},
  \bibinfo{person}{Yongfeng Huang}, {and} \bibinfo{person}{Xing Xie}.}
  \bibinfo{year}{2019}\natexlab{b}.
\newblock \showarticletitle{Npa: Neural news recommendation with personalized
  attention}. In \bibinfo{booktitle}{\emph{SIGKDD}}.
  \bibinfo{pages}{2576--2584}.
\newblock


\bibitem[\protect\citeauthoryear{Wu, Wu, Cao, Huang, and Xie}{Wu
  et~al\mbox{.}}{2021a}]%
        {wu2021fedgnn}
\bibfield{author}{\bibinfo{person}{Chuhan Wu}, \bibinfo{person}{Fangzhao Wu},
  \bibinfo{person}{Yang Cao}, \bibinfo{person}{Yongfeng Huang}, {and}
  \bibinfo{person}{Xing Xie}.} \bibinfo{year}{2021}\natexlab{a}.
\newblock \showarticletitle{Fedgnn: Federated graph neural network for
  privacy-preserving recommendation}.
\newblock \bibinfo{journal}{\emph{arXiv preprint arXiv:2102.04925}}
  (\bibinfo{year}{2021}).
\newblock


\bibitem[\protect\citeauthoryear{Wu, Wu, Ge, Qi, Huang, and Xie}{Wu
  et~al\mbox{.}}{2019c}]%
        {wu2019neural2}
\bibfield{author}{\bibinfo{person}{Chuhan Wu}, \bibinfo{person}{Fangzhao Wu},
  \bibinfo{person}{Suyu Ge}, \bibinfo{person}{Tao Qi},
  \bibinfo{person}{Yongfeng Huang}, {and} \bibinfo{person}{Xing Xie}.}
  \bibinfo{year}{2019}\natexlab{c}.
\newblock \showarticletitle{Neural news recommendation with multi-head
  self-attention}. In \bibinfo{booktitle}{\emph{EMNLP-IJCNLP}}.
  \bibinfo{pages}{6389--6394}.
\newblock


\bibitem[\protect\citeauthoryear{Wu, Wu, Qi, and Huang}{Wu
  et~al\mbox{.}}{2021b}]%
        {wu2021empowering}
\bibfield{author}{\bibinfo{person}{Chuhan Wu}, \bibinfo{person}{Fangzhao Wu},
  \bibinfo{person}{Tao Qi}, {and} \bibinfo{person}{Yongfeng Huang}.}
  \bibinfo{year}{2021}\natexlab{b}.
\newblock \showarticletitle{Empowering News Recommendation with Pre-trained
  Language Models}.
\newblock \bibinfo{journal}{\emph{arXiv preprint arXiv:2104.07413}}
  (\bibinfo{year}{2021}).
\newblock


\bibitem[\protect\citeauthoryear{Wu, Wu, Qi, and Huang}{Wu
  et~al\mbox{.}}{2021c}]%
        {wu2021two}
\bibfield{author}{\bibinfo{person}{Chuhan Wu}, \bibinfo{person}{Fangzhao Wu},
  \bibinfo{person}{Tao Qi}, {and} \bibinfo{person}{Yongfeng Huang}.}
  \bibinfo{year}{2021}\natexlab{c}.
\newblock \showarticletitle{Two Birds with One Stone: Unified Model Learning
  for Both Recall and Ranking in News Recommendation}.
\newblock \bibinfo{journal}{\emph{arXiv preprint arXiv:2104.07404}}
  (\bibinfo{year}{2021}).
\newblock


\bibitem[\protect\citeauthoryear{Wu, Qiao, Chen, Wu, Qi, Lian, Liu, Xie, Gao,
  Wu, et~al\mbox{.}}{Wu et~al\mbox{.}}{2020}]%
        {wu2020mind}
\bibfield{author}{\bibinfo{person}{Fangzhao Wu}, \bibinfo{person}{Ying Qiao},
  \bibinfo{person}{Jiun-Hung Chen}, \bibinfo{person}{Chuhan Wu},
  \bibinfo{person}{Tao Qi}, \bibinfo{person}{Jianxun Lian},
  \bibinfo{person}{Danyang Liu}, \bibinfo{person}{Xing Xie},
  \bibinfo{person}{Jianfeng Gao}, \bibinfo{person}{Winnie Wu}, {et~al\mbox{.}}}
  \bibinfo{year}{2020}\natexlab{}.
\newblock \showarticletitle{Mind: A large-scale dataset for news
  recommendation}. In \bibinfo{booktitle}{\emph{ACL}}.
  \bibinfo{pages}{3597--3606}.
\newblock


\bibitem[\protect\citeauthoryear{Zhang, Chen, Xie, Chen, Zhang, and
  Xiang}{Zhang et~al\mbox{.}}{2021}]%
        {zhang2021privacy}
\bibfield{author}{\bibinfo{person}{Xiaoyu Zhang}, \bibinfo{person}{Chao Chen},
  \bibinfo{person}{Yi Xie}, \bibinfo{person}{Xiaofeng Chen},
  \bibinfo{person}{Jun Zhang}, {and} \bibinfo{person}{Yang Xiang}.}
  \bibinfo{year}{2021}\natexlab{}.
\newblock \showarticletitle{Privacy Inference Attacks and Defenses in
  Cloud-based Deep Neural Network: A Survey}.
\newblock \bibinfo{journal}{\emph{arXiv preprint arXiv:2105.06300}}
  (\bibinfo{year}{2021}).
\newblock


\bibitem[\protect\citeauthoryear{Zhou, Zhu, Song, Fan, Zhu, Ma, Yan, Jin, Li,
  and Gai}{Zhou et~al\mbox{.}}{2018}]%
        {zhou2018deep}
\bibfield{author}{\bibinfo{person}{Guorui Zhou}, \bibinfo{person}{Xiaoqiang
  Zhu}, \bibinfo{person}{Chenru Song}, \bibinfo{person}{Ying Fan},
  \bibinfo{person}{Han Zhu}, \bibinfo{person}{Xiao Ma},
  \bibinfo{person}{Yanghui Yan}, \bibinfo{person}{Junqi Jin},
  \bibinfo{person}{Han Li}, {and} \bibinfo{person}{Kun Gai}.}
  \bibinfo{year}{2018}\natexlab{}.
\newblock \showarticletitle{Deep interest network for click-through rate
  prediction}. In \bibinfo{booktitle}{\emph{SIGKDD}}.
  \bibinfo{pages}{1059--1068}.
\newblock


\bibitem[\protect\citeauthoryear{Zhu, Liu, and Han}{Zhu et~al\mbox{.}}{2019}]%
        {zhu2019deep}
\bibfield{author}{\bibinfo{person}{Ligeng Zhu}, \bibinfo{person}{Zhijian Liu},
  {and} \bibinfo{person}{Song Han}.} \bibinfo{year}{2019}\natexlab{}.
\newblock \showarticletitle{Deep Leakage from Gradients}.
\newblock \bibinfo{journal}{\emph{NeurIPS}}  \bibinfo{volume}{32}
  (\bibinfo{year}{2019}), \bibinfo{pages}{14774--14784}.
\newblock


\end{thebibliography}

\end{document}